\documentclass[preview]{epl2} 
\usepackage{graphicx}
\usepackage{epstopdf}
\usepackage{subfigure, amsmath}

\title{Static and dynamic behavior of multiplex networks under interlink strength variation}
\shorttitle{Static and dynamic behavior of multiplex networks} 

\author{N. Bastas \thanks{These authors contributed equally to this work} \and F. Lazaridis\footnotemark[1] \and P. Argyrakis \and M. Maragakis }
\shortauthor{N. Bastas \etal}

\institute{                    
   Department of Physics, University of Thessaloniki, 54124 Thessaloniki, Greece\\

}


\abstract{
It has recently been suggested \cite{Radicchi2013} that in a two-level multiplex network, a gradual change in the value of the``interlayer'' strength $p$ can provoke an abrupt structural transition. The critical point $p^*$ at which this happens is system-dependent. In this article, we show in a similar way as in \cite{Garrahan2014} that this is a consequence of the graph Laplacian formalism used in \cite{Radicchi2013}. We calculate the evolution of $p^{*}$ as a function of system size for ER and RR networks. We investigate the behavior of structural measures and dynamical processes of a two-level system as a function of $p$, by Monte-Carlo simulations, for simple particle diffusion and for reaction-diffusion systems. We find that as $p$ increases there is a smooth transition from two separate networks to a single one. We cannot find any abrupt change in static or dynamic behavior of the underlying system.}

\begin{document}

\maketitle

\section{Introduction}
In the past several years, single networks have been extensively studied \cite{Watts1998,Barabasi1999,Barabasi2002,Mendes2002,Newman2003,Caldarelli2012} both regarding their structure and also regarding different interactions between their nodes. In real world systems, however, there may be more than one type of relationship for the same collection of objects constituting the network. Consider, for example, the communication and power networks in a given geographical region. In this case, a failure in some power station will affect not only the functionality of the power grid, but also the routing system for all computers that uses electrical power to sustain its functionality. Thus, more recently \cite{Buldyrev2010}, effort has been applied to the so called ``interdependent'' or ``interconnected'' networks, meaning a system of two or more networks that are linked together. Several articles have been published investigating the properties of these systems \cite{Buldyrev2010,Thurner2010} as well as the evolution of dynamical processes on them\cite{Mendiola2012,Gomez2013}. In their simplest form, they consist of two single networks having their nodes connected in a one-to-one configuration with each ``interlink'' having the same strength $p$ (in the range of $0<p< \infty$). A first question of interest is how does the variation of the parameter $p$ affect the global structural properties of the entire system. To this end, one can make use of the Laplacian matrix. According to graph theory, given a graph of $N$ nodes, the adjacency matrix $A$ is defined as:

\begin{equation}
\centering 
 \mathcal{L}=D-A
\end{equation}

where $D$ is a diagonal $N \times N$ matrix with the matrix elements $d_{ii}$ equal to the vertex degree. The lowest eigenvalue of this matrix $\lambda_1$ is equal $0$ in the case where the graph is connected. The other eigenvalues contribute to the dynamical processes taking part on such a network (e.g diffusion, see \cite{NewmanBook2010}). The most important role is played by the second smallest eigenvalue of the Laplacian, $\lambda_2$, which is known as algebraic connectivity \cite{Fiedler1973}. 

Recently, the authors in \cite{Radicchi2013} have used this technique to investigate how the strength of the ``interlinks'' affects the structural properties of a two-level multiplex network. A multiplex network is a special case of interdependent networks, where nodes at each layer are instances of the same entity. The configuration they used consisted of two coupled undirected networks each with the same number of nodes $N$. Nodes in the fist network were connected one-to-one with the nodes of the second network by links of equal strength $p$. Thus, the number of interconnected links is also $N$. Then, the so called ``supra-Laplacian'' matrix acquires the form:

\begin{equation}
\label{eq:supraLaplacian}
\mathcal{L} = \left( \begin{array}{cc}
\mathcal{L}_A + pI_N & -pI_N \\
-pI_N & \mathcal{L}_B + pI_N \\
\end{array} \right)
\end{equation}  

where $I_N$ is the identity matrix of dimension $N \times N$, $\mathcal{L}_A$ and $\mathcal{L}_B$ are the Laplacians of networks $A$ and $B$, respectively. The blocks are symmetric matrices of dimension $N \times N$ (actually this procedure is equivalent to taking the Laplacian matrix of the overall multiplex, but now it is in a form which helps to understand the contribution of $p$). 

The main finding of ref. \cite{Radicchi2013} was that there is a critical value of $p$, $p^{*}$, for which the ``algebraic connectivity'' ($\lambda_2$) of the system changes abruptly. For values of $p<p^{*}$, the system behaves as two separate networks, while for $p>p^{*}$ the system behaves like one single network. This happens at a value of $p$ for which $d\lambda_2/dp$ is discontinuous. This discontinuity comes as a result of a crossing that occurs between the eigenvalues of the supra-Laplacian. In plain terms, above $p^{*}$ the ranking of the eigenvalues changes. Thus, the eigenvalue that is initially the second smallest for $p < p^{*}$, $\lambda_2$, becomes larger than another eigenvalue ($\lambda_i$) above $p > p^{*}$. This is manifested by a discontinuity in the derivative of $\lambda_2$ at $p^{*}$. 

Very recently, Garrahan and Lesanovsky \cite{Garrahan2014} have raised objections at the interpretation of the findings in \cite{Radicchi2013}. They pointed out that the eigenvalue crossing is only a consequence of the reducibility of the matrix and they applied it for the case of two identical multiplexed networks. The characteristic polynomial of the supra-Laplacian matrix $\mathcal{L}$ (eq.(\ref{eq:supraLaplacian})) is given by: 

\begin{equation}
\left| \begin{array}{cc}
\centering
 \mathcal{L}_A+pI_N - \lambda^{\mathcal{L}} I_N & -pI_N \\
-pI_N & \mathcal{L}_B+pI_N- \lambda^{\mathcal{L}} I_N 
\end{array} \right| = 0 
\end{equation}

If $\mathcal{L}_A=\mathcal{L}_B$, then:

\begin{equation}
\centering 
\left| \mathcal{L}_A -\lambda^{\mathcal{L}} I_N \right|\left| \mathcal{L}_A - (\lambda^{\mathcal{L}}-2p) I_N\right|=0
\label{eq3}
\end{equation}

From eq.(\ref{eq3}), either $\left|\mathcal{L}_A -\lambda^{\mathcal{L}} I_N\right| $ or $\left|\mathcal{L}_A - (\lambda^{\mathcal{L}}-2p)I_N\right|$ should be equal to $0$. For the former, we get: $\lambda^{\mathcal{L}}_i=\lambda^{\mathcal{L}_A}_i$ and for the latter $\lambda^{\mathcal{L}}_i=\lambda^{\mathcal{L}_A}_i+2p$. For a connected system (such as network $A$), $0=\lambda_1^{\mathcal{L}_A} \leq \lambda_2^{\mathcal{L}_A} ... \leq \lambda_N^{\mathcal{L}_A}$. Thus, depending on the value of $p$, we may have the following two cases: $\lambda^{\mathcal{L}}_1=0$,$\lambda^{\mathcal{L}}_2=\lambda_2^{\mathcal{L}_A}$, $\lambda^{\mathcal{L}}_3=2p$ or $\lambda^{\mathcal{L}}_1=0$,$\lambda^{\mathcal{L}}_2=2p$, $\lambda^{\mathcal{L}}_3=\lambda_2^{\mathcal{L}_A}$. Thus, for a suitable values of $\lambda_2^{\mathcal{L}_A}$ and $p$, we will have an interchange in the relative position of $\lambda_2^{\mathcal{L}}$ and $\lambda_3^{\mathcal{L}}$ and the crossing observed in \cite{Radicchi2013}.

In the present work, we investigate the effect of the parameter $p$ on various topological and dynamical properties of a two-layered multiplex network. Specifically, we want to uncover the nature of the $p^{*}$ transition, i.e. is it a sharp, sudden transition or is it a smooth one. We calculate numerically the ``algebraic connectivity'' with respect to interlink strength for ER and RR networks of various sizes using standard simulation techniques. We investigate the evolution of static structural properties (mean shortest path) and then we proceed with various dynamic processes (simple diffusion and reaction-diffusion) to see if we can infer the nature of any structural changes. Furthermore, we investigate the behavior of algebraic connectivity as a function of system size. We finish by presenting our main conclusions.

\section{Method}

We used two coupled networks of the same size $N$ and the same statistical properties, in a multiplex configuration. The specifications of each configuration are defined in the respective figure captions. In order to specify the dynamical properties of the systems, we perform simple particle diffusion for a collection of particles with excluded volume, and reaction-diffusion processes of two types of particles ($A+B$ model). These are performed by standard Monte-Carlo techniques using random walks \cite{Argyrakis1992}. For both processes, we have two types of particles, $A$ and $B$, initially placed in the two different layers, one kind on each layer, with a prescribed density $\rho$. We let the particles to diffuse. For the simple diffusion, we calculate the cumulative number of collisions $m$ between the different types of particles. We estimate the value $\tau_{col}$, which is the time when a certain total number of collisions between different types of particles has occurred. For the reaction-diffusion process, where steady state condition is imposed, we calculate the mean time required for $\rho N$ reactions to occur. In this model, $A$ reacts with $B$, but not with another $A$, and similarly for the $B$ particles. We should point out that when two particles react they are removed from the system and two new particles (an $A$ and a $B$) are randomly placed on the layer that corresponds to their type.

We also use the exact enumeration calculation for the case of simple diffusion \cite{Havlin2004}. In the case of a random regular (RR) network, the system is symmetric. Thus, the following set of mean-field recursive equations govern the dynamics of the system:

\begin{equation}
\begin{aligned}
f_{L1}(t+1)=\frac{k}{k+p}f_{L1}(t)+\frac{p}{k+p}f_{L2}(t) \\
f_{L2}(t+1)=\frac{k}{k+p}f_{L2}(t)+\frac{p}{k+p}f_{L1}(t)
\end{aligned}
\label{eqRR}
\end{equation}

where $f_{L1}(t)$ is the state of every node on layer $L1$ and $f_{L2}(t)$ the state of every node on layer $L2$. In each step $t$, a node contributes a fraction of $\frac{1}{k+p}f_{i}(t)$ to each one of its $k$ neighbors on layer $i$ and $\frac{p}{k+p}f_{i}(t)$ to the corresponding node on the other layer. We specify the time to reach equilibrium (denoted as $\tau_{enum}$) by solving for initial conditions $f_{L1}(0)=1$ and $f_{L2}(0)=0$, and using the stopping condition $\frac{p}{p+k}(f_{L1}(t)-f_{L2}(t)) < 10^{-6}$.

The situation is more complicated for the case of ER networks, since they do not have a symmetric topology. We consider the mirror case (identical layers), and focus on a node with degree $k$. Denote with $a_k(t)$ the containment of this node and with $b_k(t)$ the containment of its corresponding node (which must have the same degree $k$). A node with degree $k$ has on average $\frac{k'P(k')}{\left\langle k \right\rangle}$ nodes with degree $k'$ and from each node a fraction of $a_{k'}(t)\frac{1}{k'+p}$ is passed to the node with degree $k$. The overall contribution should be $k \sum_{k'=1}^{+\infty}{\frac{k'P(k')a_{k'}(t)}{k'+p}}$. Also, a fraction $\frac{p}{k+p}b_k(t)$ will be received from the other layer. The same holds for the nodes on the other layer. Thus, the following recursive equations hold:

\begin{widetext}
\begin{equation}
\begin{aligned}
a_k(t+1)=\frac{k}{\left\langle k \right\rangle} \sum_{k'=1}^{+\infty}{a_{k'}(t)P(k')} -  \frac{pk}{\left\langle k \right\rangle} \sum_{k'=1}^{+\infty}{\frac{a_{k'}(t)}{k'+p}P(k')} +  \frac{p}{k+p}b_k(t) \\
b_k(t+1)=\frac{k}{\left\langle k \right\rangle} \sum_{k'=1}^{+\infty}{b_{k'}(t)P(k')} - \frac{pk}{\left\langle k \right\rangle} \sum_{k'=1}^{+\infty}{\frac{b_{k'}(t)}{k'+p}P(k')} + \frac{p}{k+p}a_k(t) 
\end{aligned}
\label{eqER}
\end{equation}
\end{widetext}

We solve eq. (\ref{eqER}) for initial conditions $a_k(0)=1$ and $b_k(0)=0$, and using the stopping condition $\sum_{k=1}^{kmax}{\left|a_k(t)-b_k(t)\right|\frac{p}{k+p} P(k)}< 10^{-6}$. Thus, we acquire the time for the system to reach the equilibrium, $\tau_{enum}$.

Furthermore, the change of the structural properties of the system as a function of $p$ is investigated using the average shortest path ($\left\langle l_{sp} \right\rangle$) by varying the interlink strength $p$ from nearly zero to $100$, using a modified version of the Brandes algorithm \cite{Brandes2001} for weighted networks. We finally calculate the evolution of $p^{*}$ as a function of $N$, using the methodology proposed in \cite{Radicchi2013}.

\section{Results and Discussion}

\begin{figure}[h!]
\centering 
\includegraphics[width=8cm] {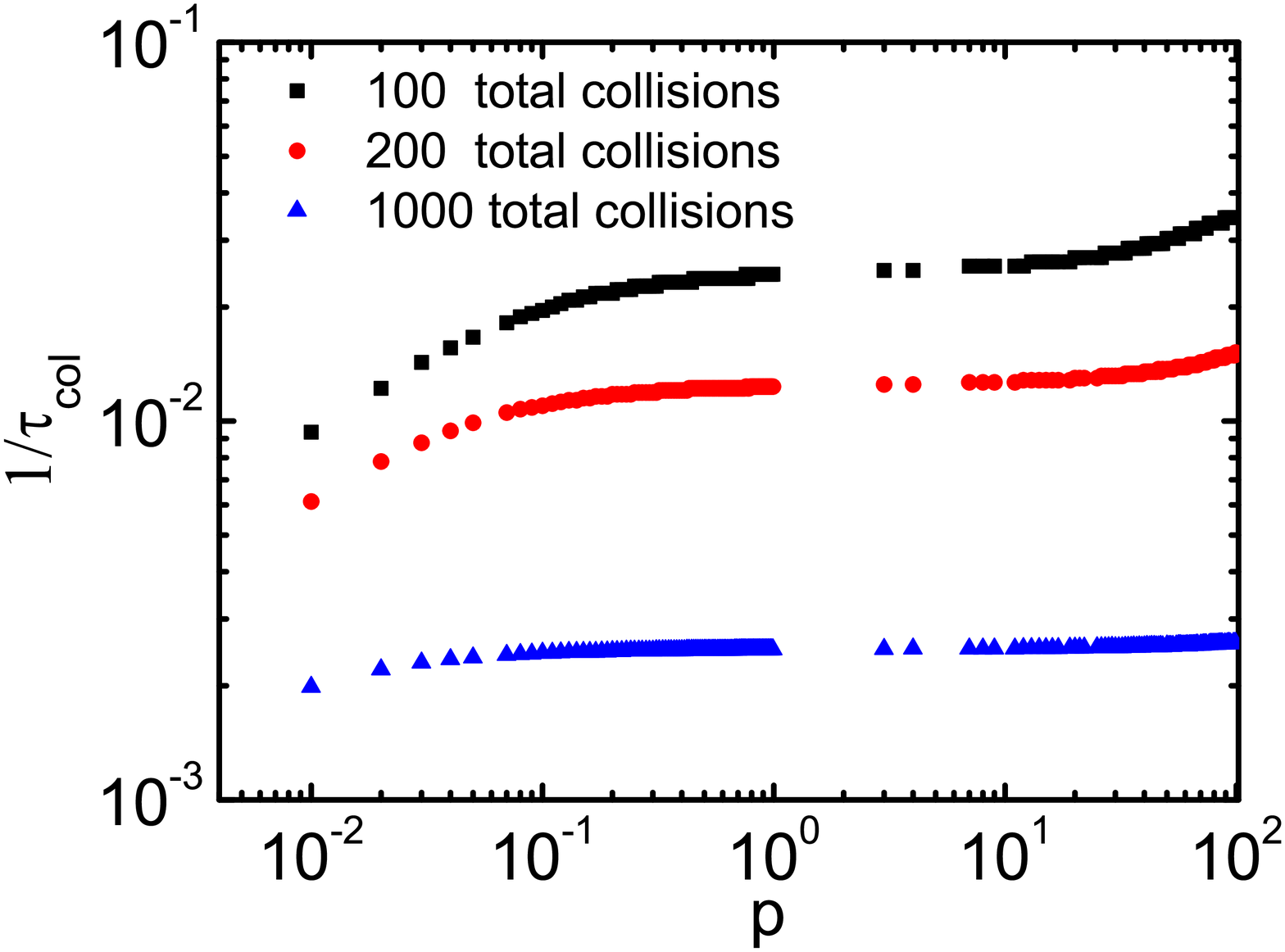}
\caption{(Color Online). (a) Plot of the inverse of the time, $1/\tau_{col}$, at which $m$ total collisions between $A$ and $B$ particles have been made as a function of $p$, on RR networks with $k=4$, $N=1000$ and $\rho_A=\rho_B=0.05$, averaged over $100$ realizations. The plot is in $log-log$ scale. Data points are: $m=100$ (black squares), $m=200$ (red circles) and $m=1000$ (blue triangles). For all the observed cases, the transition is smooth as $p$ increases.} 
\label{Figure1}
\end{figure}

We first examine the evolution of $1/\tau_{col}$ as a function of $p$, for different cutoff values of $m$ ($m$ being the cumulative number of collisions between the different types of particles). In Fig. \ref{Figure1}, we present the results for two multiplexed RR networks with $k=4$, $N=1000$ and $\rho_A=\rho_B=0.05$, averaged over $100$ realizations. We observe that the quantity $1/\tau_{col}$ changes smoothly with $p$. Thus, there is no indication that any structural transition can manifest itself through this process. 

\begin{figure}[!h]
\centering 
\includegraphics[width=8cm] {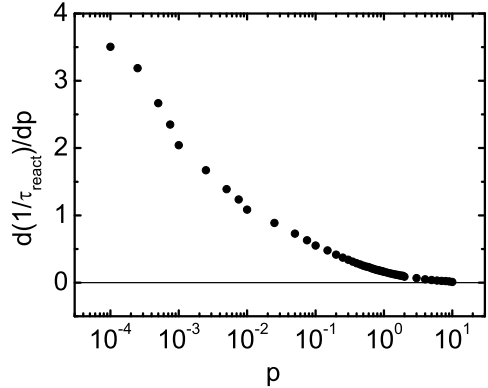} 
\caption{Plot of the derivative of the inverse characteristic time ($1/\tau_{react}$) as a function of $p$, for two identical ER networks with $\left\langle k \right\rangle = 5$ and $N=1000$.} 
\label{Figure2}
\end{figure}

We next perform the $A+B \rightarrow 0$ reaction-diffusion process and calculate the mean time required for all particles to react, $\tau_{react}$. The density of particles $A$ and $B$ is exactly the same ($\rho = 0.25$) and remains fixed throughout the process (steady state). All $A$ particles are initially placed on one layer and all $B$ particles on the other. We calculate the time $\tau_{react}$ which is the time it takes to have $\rho N$ reactions. In Fig. \ref{Figure2}, we plot the first derivative of $1/\tau_{react}$ as a function of $p$. As $p$ increases, this quantity tends to $0$, meaning that $\tau_{react}$ is independent of $p$ for large $p$ values. If the system were to undergo an abrupt structural transition for a specific value of $p$, we should have observed qualitatively this sharp change in the vicinity of such a point. 

\begin{figure}[!]
\centering 
\subfigure[]{
\includegraphics[width=8cm] {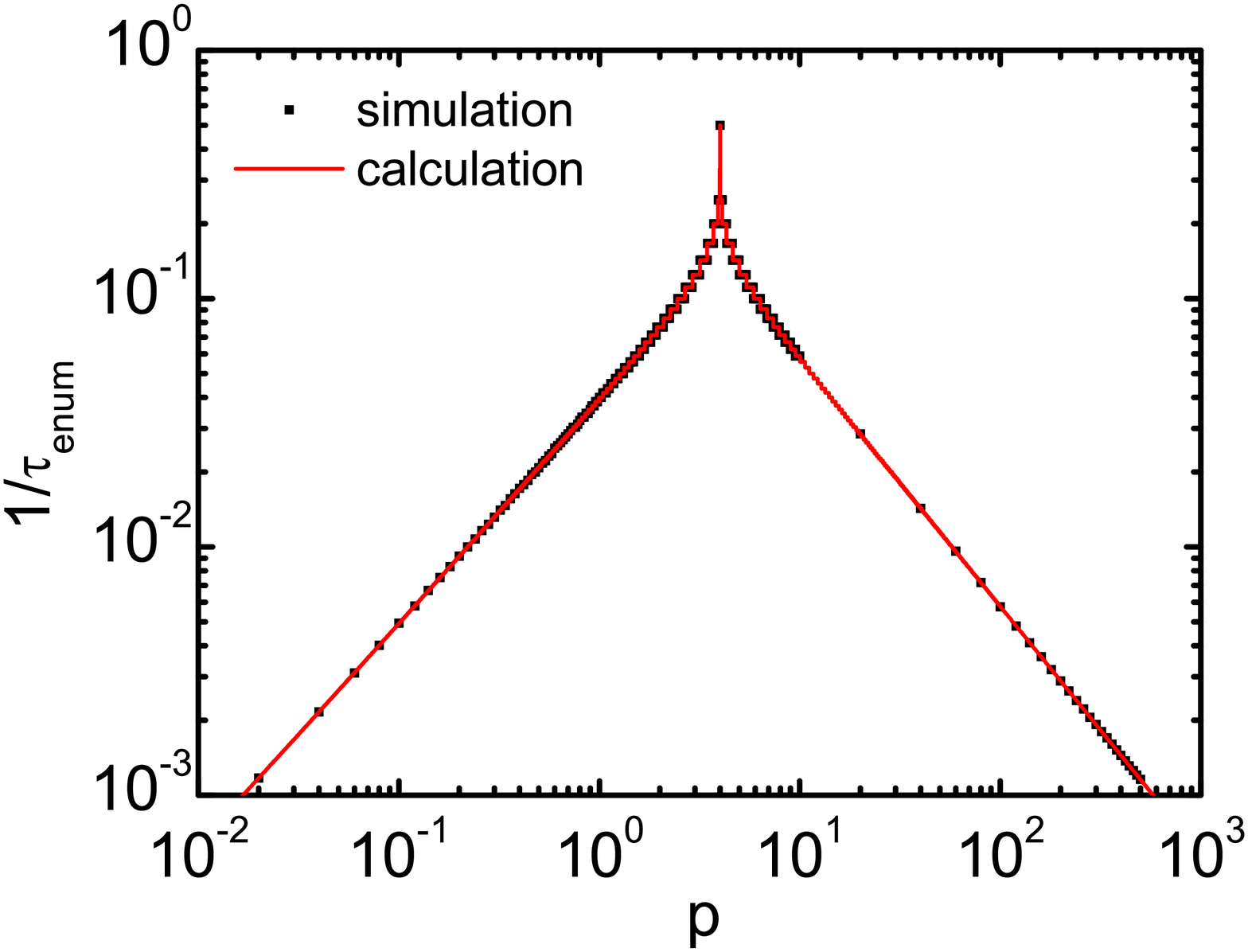} \label{Figure3a} }
\subfigure[]{
\includegraphics[width=8cm] {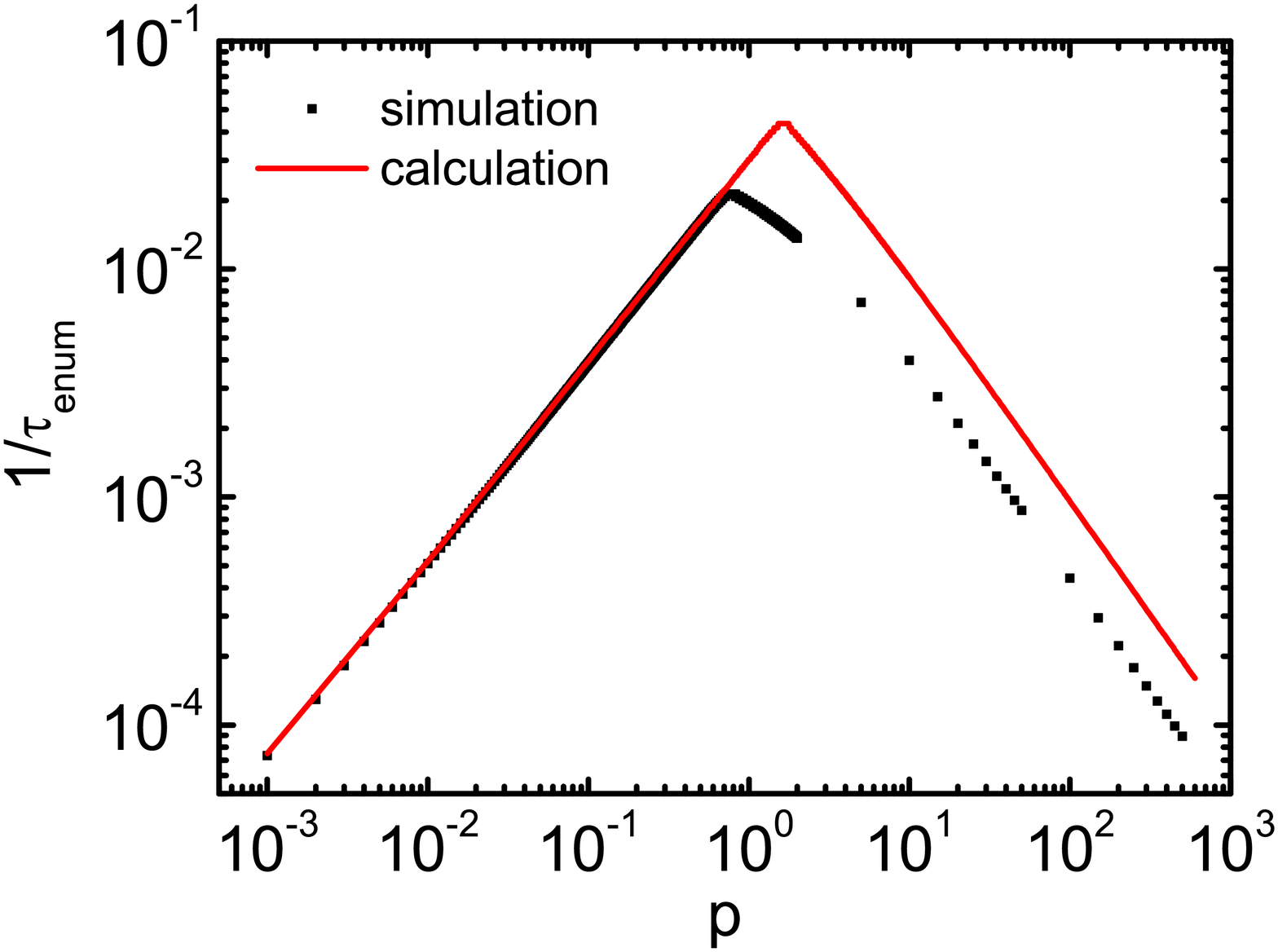} \label{Figure3b} }
\caption{(Color Online). Plot of the inverse of the equilibration time ($1/\tau_{enum}$) as a function of $p$, for the case of (a) two coupled random-regular networks with $k=4$ and $N=1000$ and (b) two coupled identical ER networks with $\left\langle k \right\rangle = 5$ and $N=1000$. In (a), there is a peak at $p \simeq 4$. In both figures, there is a change in the behavior for a value of $p$. However the point at which this occurs does not coincide with the $p^*$ calculated with the method proposed in \cite{Radicchi2013}.} 
\label{Figure3}
\end{figure}

Finally, we use the exact enumeration method, for the case of the diffusion process. In Fig. \ref{Figure3}, we plot the inverse of the equilibration time ($1/\tau_{enum}$) as a function of $p$ for a setting of two identical multiplexed RR networks ($k=4$, $N=1000$) and ER networks ($\left\langle k \right\rangle=5,N=1000$). In Fig \ref{Figure3a}, a peak appears at $p \simeq 4$, which is not the value calculated with the method used in \cite{Radicchi2013}, which was $p^{*} \simeq 0.41$. The same behavior is recovered when $k=8$ for RR networks (data not shown). The results are accompanied by the numerical simulation of the process, and as we see they are in excellent agreement. A qualitative explanation can be given here: when $p < k$, it happens that $p/k+p < k/k+p$, and the majority of particles remain for longer times in the starting layer. As $p$ increases particles tend to move to the other layer, thus promoting the decrease of the equilibration time. When $p > k$, we have that $p/k+p > k/k+p$, and thus the inverse procedure takes place. The case is similar for coupled ER networks, see Fig. \ref{Figure3b}, except for the fact that the peak is not sharp since in this case there is an inherent asymmetry in the degree distribution. 

\begin{figure}[!]
\centering 
\includegraphics[width=8cm] {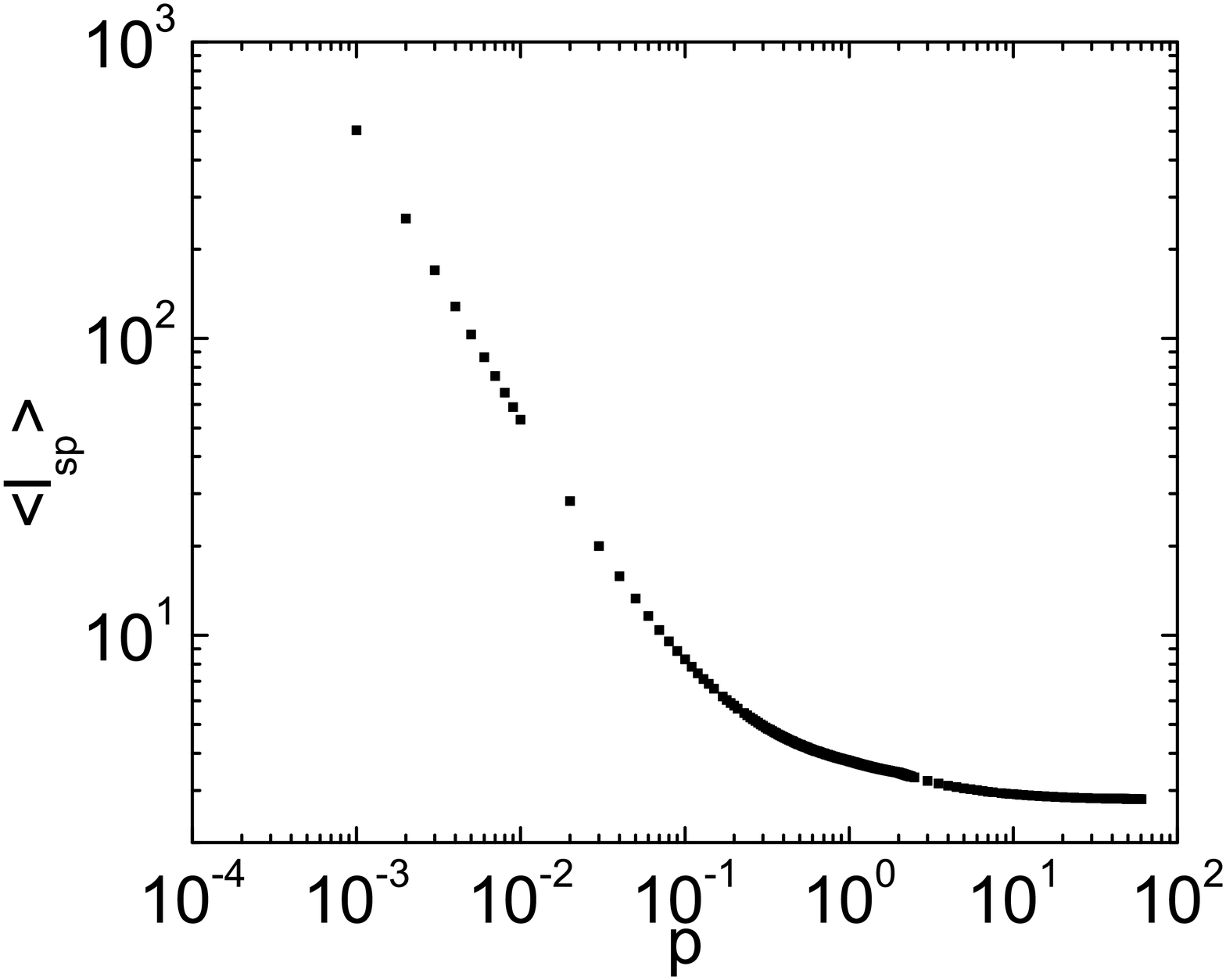}
\caption{Plot of the ``average shortest path length''  ($\left\langle l_{sp} \right\rangle$)  as a function of $p$ for a multiplex of two random regular networks with $k=8$ and $N=1000$ each. It is obvious that there is no abrupt transition of this quantity.} 
\label{Figure4}
\end{figure}

These results indicate that there is no manifestation of any sudden structural change on common dynamical processes on multiplex networks. We further proceed by investigating network properties of the multiplex systems. We search to see if there is some structural modification in the topological properties of the network by investigating the ``average shortest path'' ($\left\langle l_{sp} \right\rangle$). We performed calculations on a random-regular multiplex network with $k=8$ and $N=1000$, using the algorithm proposed in \cite{Brandes2001} for the case of weighted graphs. Regarding the weights, we used a value of 1 for all intralayer connections and $p$ for the interlayer ones. We define the distance between two nodes connected by an interlink as the inverse of its strength $p$. The results for $\left\langle l_{sp} \right\rangle$ are shown in Fig. \ref{Figure4}, where we observe that as $p \rightarrow 0$, $\left\langle l_{sp} \right\rangle \rightarrow \infty$ because $1/p \rightarrow \infty$. Furthermore, when $p \ll 1$, $\left\langle l_{sp} \right\rangle$ is controlled by the maximum shortest path between a source node in network A(B) and a target node in network B(A). As $p$ increases, $1/p$ decreases and so does $\left\langle l_{sp} \right\rangle$. For $p \gg 1$, the multiplex behaves as a single network. We observe that, by varying the value of $p$, this quantity changes in a continuous way. Thus, no abrupt transition can be seen here from the data as well.

\begin{figure}[!h]
\centering 
\subfigure[]{
\includegraphics[width=8cm] {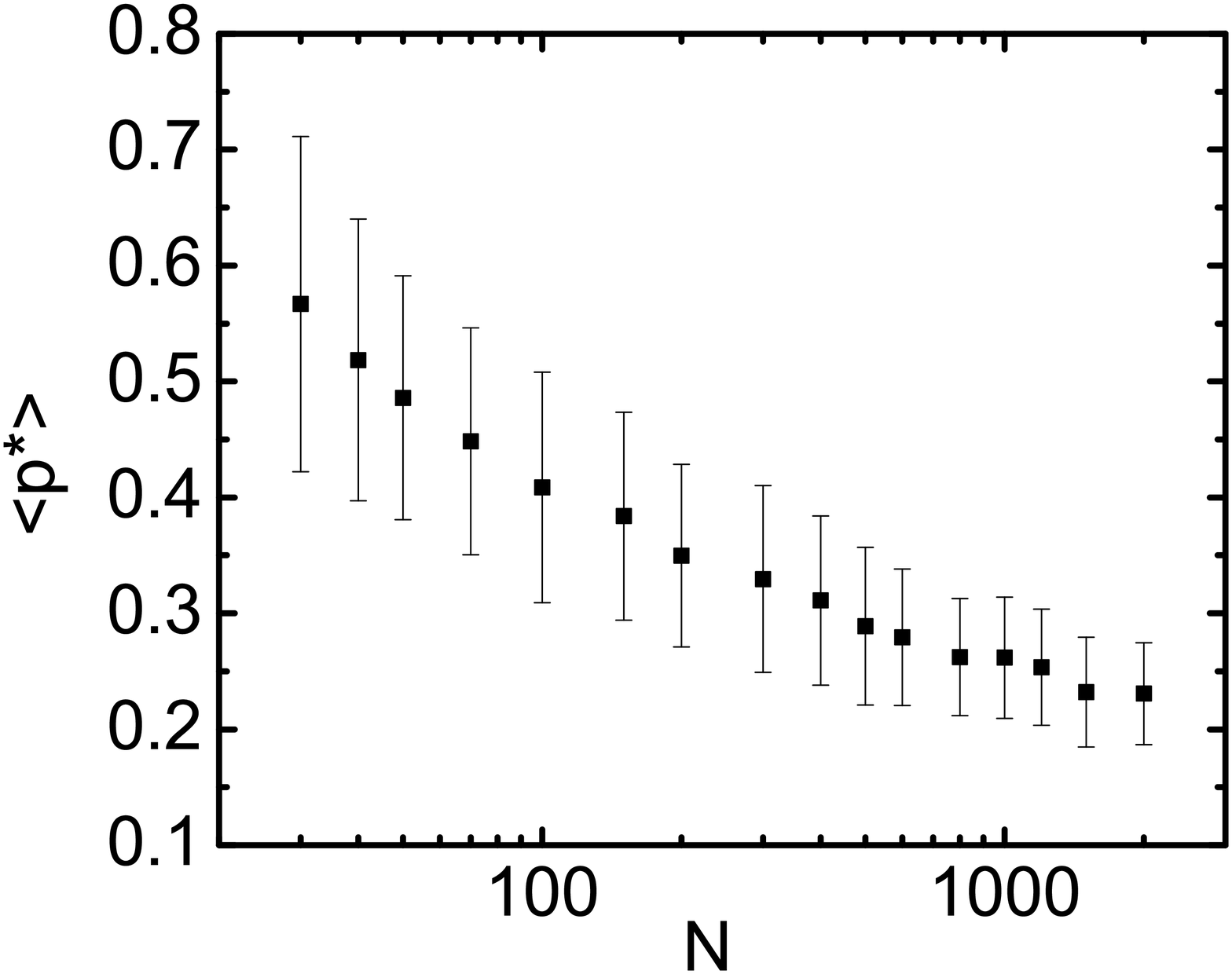} \label{Figure5a} }
\subfigure[]{
\includegraphics[width=8cm] {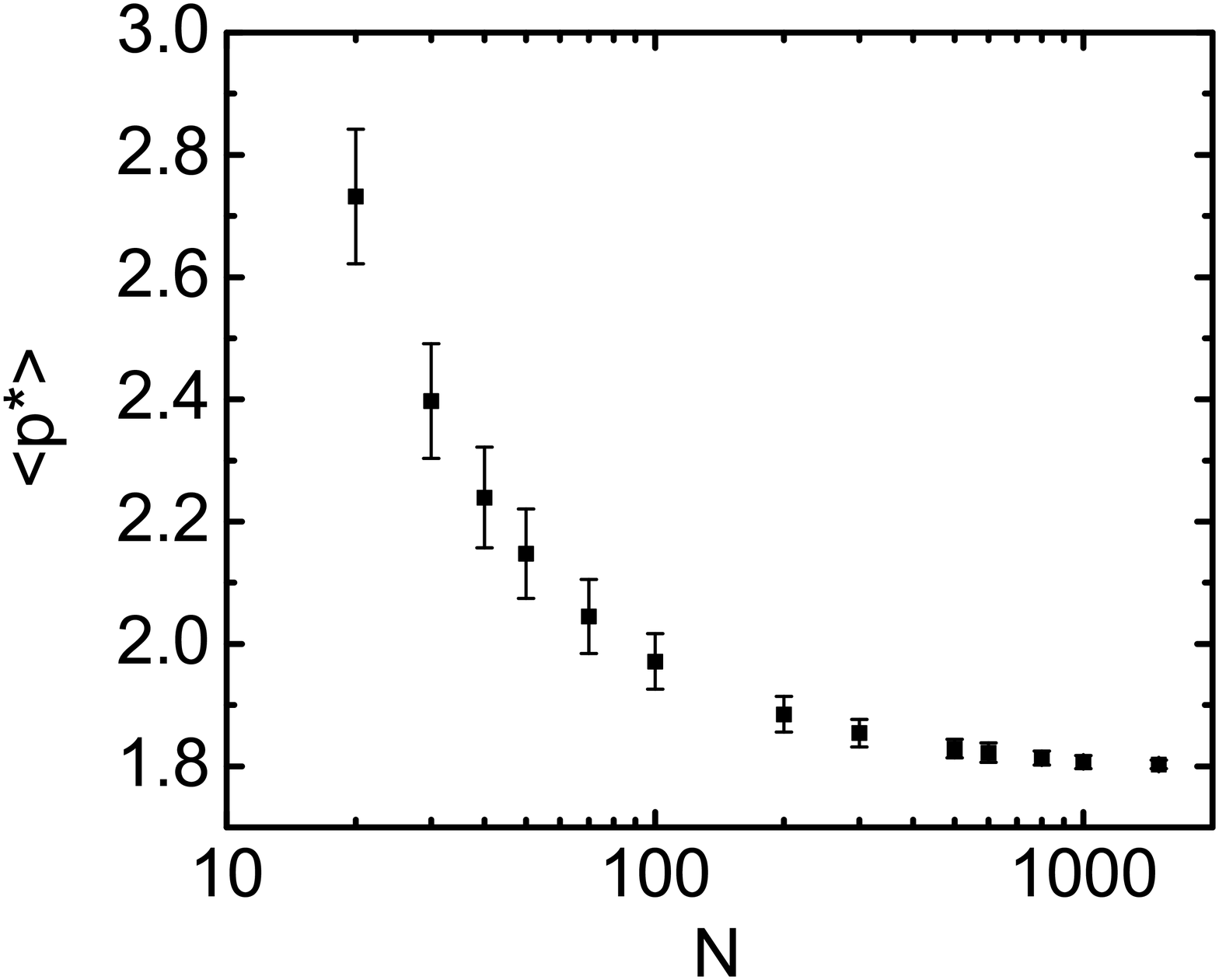}  \label{Figure5b} }
\caption{Plot of the value of the transition point of the algebraic connectivity, $p^{*}$, as a function of the size of each layer $N$, on (a) two coupled ER networks with $\left\langle k \right\rangle = 5$ and (b) two coupled RR networks with $k = 8$. Data for both cases are averaged over $100$ runs for large systems and up to $5000$ for small ones. It is evident that $p^{*}$ is system and size dependent.} 
\label{Figure5}
\end{figure}

We now apply the methodology proposed in \cite{Radicchi2013} (see also eq. (\ref{eq:supraLaplacian})) to investigate the behavior of the algebraic connectivity, $\lambda_2$, as a function of the size of the layers. In Fig. \ref{Figure5}, we plot the evolution of $p^{*}$ as a function of system size $N$ for two types of networks:  ER networks with $\left\langle k \right\rangle = 5$ (Fig.\ref{Figure5a}) and random regular with $k=8$ (Fig.\ref{Figure5b}). We observe a clear dependence of $p^{*}$ on the size $N$ of the layers. Also, there is a fast decay on the value of $p^{*}$, but we cannot determine its nature (i.e. being power-law or exponential), due to the small range of $N$.

\section{Conclusions}

In this paper we have tested structural as well as dynamic properties of multiplex networks. We used various types of networks such as coupled RR and ER (identical or not) networks and various different methodologies. We have observed a continuous change of the structural properties as a function of $p$, as shown by the behavior of the average shortest path ($\left\langle l_{sp} \right\rangle$). We have performed diffusion and reaction-diffusion processes on the above multiplex configurations studying suitable characteristic times associated with the structural properties of the system. In all cases we verified that changes take place in a continuous manner. Also, the point where such changes occur does not coincide with the $p^{*}$ calculated with the method proposed in \cite{Radicchi2013}.

\section{Acknowledgments}
Results presented in this work have been produced using the European Grid Infrastructure (EGI) through the National Grid Infrastructures $NGI \_ GRNET$ (HellasGrid) as part of the SEE Virtual Organisation. This research was supported by European Commission FP$7$-FET project Multiplex No. $317532$. N. B. acknowledges financial support from Public Benefit Foundation Alexander S. Onassis.


\begin{thebibliography}{0}

\bibitem{Radicchi2013}
  \Name{Radicchi F. \and Arenas A.}
  \REVIEW{Nat. Phys.}{9}{2013}{717}.

\bibitem{Garrahan2014}
  \Name{Garrahan J.P. \and Lesanovsky I.}
  \REVIEW{arXiv:1406.4706}{}{2014}{}.
  
\bibitem{Watts1998}
  \Name{Watts D.J. \and Strogatz S.H.}
  \REVIEW{Nature}{393}{2014}{440}.
  
\bibitem{Barabasi1999}
  \Name{Barabasi A.L. \and Albert R.}
  \REVIEW{Science}{286}{1999}{509}.
  
\bibitem{Barabasi2002}
  \Name{Albert R. \and Barabasi A.L.}
  \REVIEW{Rev. Mod. Phys.}{74}{2002}{47}.
  
\bibitem{Mendes2002}
  \Name{Dorogovtsev S.N. \and Mendes J.F.F.}
  \REVIEW{Adv. Phys.}{51}{2002}{1079}.
  
\bibitem{Newman2003}
  \Name{Newman M.E.J.}
  \REVIEW{SIAM Rev.}{45}{2003}{167}.
 

\bibitem{Caldarelli2012}
  \Name{Caldarelli G. \and Catanzaro M.}
  \Book{Networks: A very Short Introduction}
  \Publ{Oxford University Press}
  \Year{2012}.
  
\bibitem{Buldyrev2010}
  \Name{Buldyrev S.V., Parshani R., Paul G., Stanley H.E. \and Havlin S.}
  \REVIEW{Nature}{464}{2010}{1025}.
  
\bibitem{Thurner2010}
  \Name{Szella M., Lambiotte R. \and Thurner S.}
  \REVIEW{Proc. Natl. Acad. Sci. USA}{107}{2010}{13636}.
  
\bibitem{Mendiola2012}
  \Name{Saumell-Mendiola A., Serrano M.A. \and Boguna M.}
  \REVIEW{Phys. Rev. E}{86}{2012}{026106}.
	
\bibitem{Gomez2013}
  \Name{Gomez S., Diaz-Guilera A., Gomez-Gardenes J., Perez-Vicente C.J., Moreno Y. \and Arenas A.}
  \REVIEW{Phys. Rev. Lett.}{110}{2013}{028701}.
	
\bibitem{NewmanBook2010}
  \Name{Newman M.E.J.}
  \Book{Networks: An Introduction}
  \Publ{Oxford University Press}
  \Year{2010}.
	
\bibitem{Fiedler1973}
  \Name{Fiedler M.}
  \REVIEW{Czech. Math. J.}{23}{1973}{298}.

\bibitem{Argyrakis1992}
  \Name{Argyrakis P.}
  \REVIEW{Comput. Phys.}{6}{1992}{525}.
	
\bibitem{Havlin2004}
  \Name{Ben-Avraham D. \and Havlin S.}
  \Book{Diffusion and Reactions in Fractals and Disordered Systems}
  \Publ{Cambridge University Press}
  \Year{2004}.
	
\bibitem{Brandes2001}
  \Name{Brandes U.}
  \REVIEW{J. Math. Soc.}{25}{2001}{163}.

\end{thebibliography}
\end{document}